\documentclass[aps,preprint,prb,preprintnumbers,amsmath,amssymb]{revtex4}
\usepackage[dvips,final]{graphicx}
\usepackage{dcolumn}
\usepackage{bm}
\usepackage[usenames]{color}
\usepackage{ulem}

\begin{document}

\title{Magnetoelectric effect in organic molecular solids}
\author{
Makoto Naka and Sumio Ishihara$^{\ast}$
}
\affiliation{Department of Physics, Tohoku University, Sendai 980-8578, Japan. \\
$^\ast$e-mail: ishihara@cmpt.phys.tohoku.ac.jp}
\date{\today}
\maketitle
{\bf
The Magnetoelectric (ME) effect in solids is a prominent cross correlation phenomenon, in which the electric field (${\bm E}$) controls the magnetization (${\bm M}$) and the magnetic field (${\bm H}$) controls the electric polarization (${\bm P}$). 
A rich variety of ME effects and their potential in practical applications have been investigated so far within the transition-metal compounds. 
Here, we report a possible way to realize the ME effect in organic molecular solids, in which two molecules build a dimer unit aligned on a lattice site. 
The linear ME effect is predicted in a long-range ordered state of spins and electric dipoles, as well as in a disordered state.  
One key of the ME effect is a hidden ferroic order of the spin-charge composite object. 
We provide a new guiding principle of the ME effect in materials without transition-metal elements,  which may lead to flexible and lightweight multifunctional materials. 
}
\vfill
\eject

The coupling between electric and magnetic polarizations in insulating solids has been accepted as exciting phenomena since the Curie's early prediction of the ME effect~\cite{curie}. 
A keystone in researches on the ME effect was brought by Cr$_2$O$_3$, for which a number of experimental and theoretical results have been reported since the discovery~\cite{dzyaloshinskii, astrov, folen, smolenskii}.  
Interest in the ME effect has been recently revived~\cite{fiebig,cheong}. 
This is ascribed to the several recent developments:  
i) large non-linear ME effects discovered in TbMnO$_3$ and other multiferroic materials with spin frustration~\cite{kimura, katsura, dagotto, mostovoy, kimura2, arima, khomskii}, 
ii) significant development of synthesis techniques for artificial ME composites, e.g. BaTiO$_3$/CoFe$_2$O$_4$~\cite{srinivasan, zheng}, 
and iii) a new theoretical framework for the ME polarizability, which is related to the axion electrodynamics, 
by which the ME tensor is evaluated qualitatively from first principles~\cite{essin,malashevich}. 
Almost all of the ME materials examined so far are transition-metal compounds, containing a vast treasury of magnetic and ferroelectric phenomena, allowing a coupling between the two. 

Organic molecular solids are another class of materials, in which a wide variety of magnetic and dielectric phenomena emerges~\cite{fukuyama}. 
However, to the best of our knowledge, reports of the ME effect in organic molecular solids without transition-metal elements are limited so far. 
The fundamental unit of crystalline and electronic structures in this class of materials is the molecule rather than ion/atom in the transition-metal compounds.  
In particular, flexible $\pi$ molecular orbitals prescribe their magnetic and dielectric responses. 
Low-dimensional organic molecular solids, in which the molecular dimer units build a framework of the crystal lattice, have been ubiquitously targeted as multifunctional materials in recent decades. 
Series of tetramethyl-tetrathiafulvalene (TMTTF) and bis(ethylenedithio)tetrathiafulvalene (BEDT-TTF) compounds are the well-known examples. 
A rich variety of phenomena, e.g., superconductivity, quantum spin liquid state, ferroelectricity, and so on, have attracted considerable interest, and these have been ascribed to the molecular orbitals (MO) in dimer units~\cite{kanoda, jawad, lang, iguchi, dressel, yakushi}. 

Here, we show that this dimer-type organic molecular solids provide an appropriate framework for the ME effect. 
We first present a symmetrical consideration for the ME effect in a simple one-dimensional chain model. 
Numerical calculations in a two-dimensional lattice modeling the $\kappa$-(BEDT-TTF) type organic molecular solids demonstrate that the linear ME effect emerges in a long-range ordered state of spins and electric dipoles owing to the electronic degree of freedom inside the molecular dimers. 
We identify that the essence of this phenomenon is attributable to a hidden ferroic order of the spin-charge composite object. 
The ME effect is also observed even in the spin and charge disordered state, in which the spin-charge composite ferroic order is realized. 
The present study of the ME effect provides a new strategy of material designs for a new type of multiferroic organic molecular solids. 

\begin{figure} 
\begin{center}
  \includegraphics[width=0.9\columnwidth,clip]{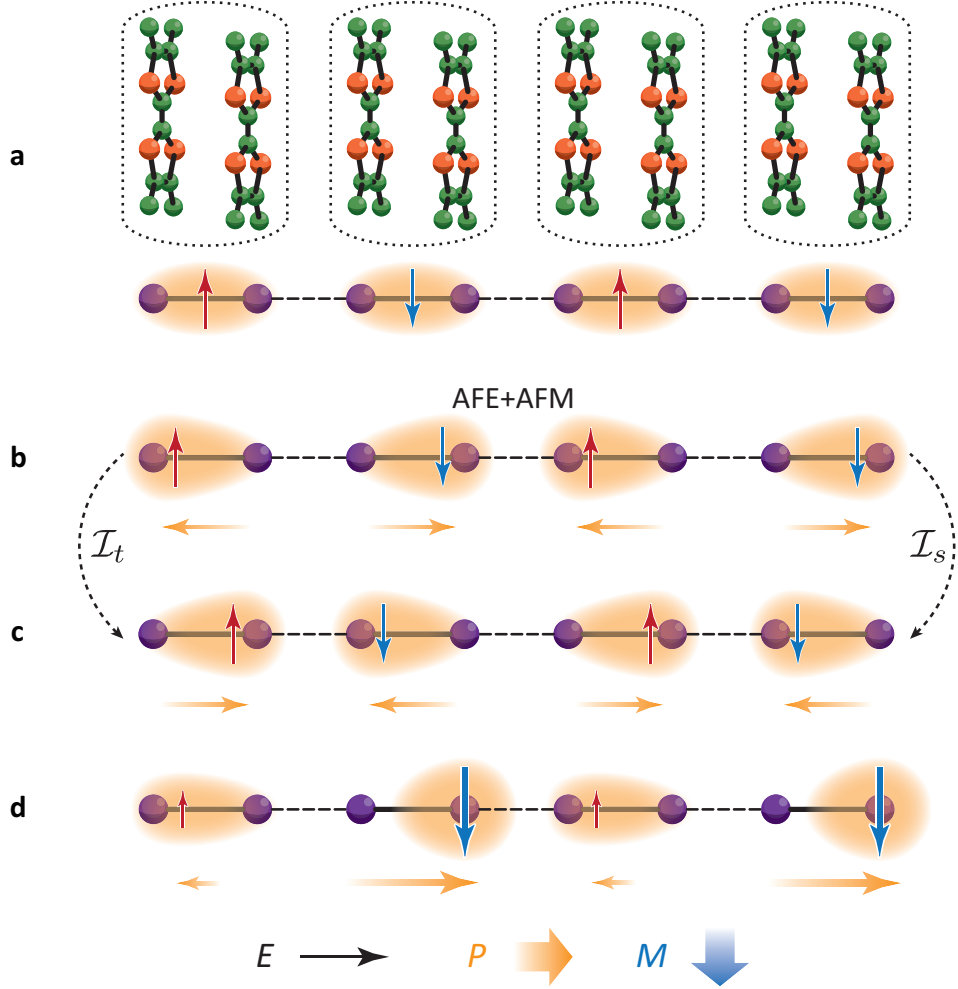}
\end{center}
\caption
{
{\bf Spin and charge configurations in a molecular dimer system where dimer units are arrayed on a one dimensional chain.} 
Filled purple circles, shaded ovals and thin red and blue arrows represent molecules, charge distributions, and spins, respectively. 
Shaded orange arrows represent electric dipole moments due to deformations of the charge clouds. 
Spins are aligned antiferromagnetically and are directed perpendicular to a chain. 
${\bm a}$, An electrically non-polarized state in which charge distributions in two molecules inside a dimer are equivalent. 
A corresponding one-dimensional chain composed of the TMTTF molecular dimers are presented in the upper panel where dotted circles denote the dimer units. 
${\bm b}$, An electrically polarized state in which charge distributions are polarized alternately, termed the multiferroic AFE+AFM state. 
${\bm c}$, A spin and charge configuration obtained by applying the time reversal operation (${\cal I}_{t}$) or the space inversion operation (${\cal I}_{s}$) to the configuration in ${\bm b}$. 
${\bm d}$, Electric field ($E$) along the chain direction is applied on the polarized state in ${\bm b}$. 
Net magnetization ($M$) and electric polarization ($P$) are induced to be perpendicular and parallel to a chain, respectively. 
}
\label{fig:one}
\end{figure}

Let us start from a simple example, a one-dimensional array of the molecular dimer units, as shown in Fig.~\ref{fig:one}, where the number of electrons are fixed to be equal to the number of the dimer units.  
This is a model for the one-dimensional chain in (TMTTF)$_2$X (X: monovalent anion). 
When a coupling between the two molecules in a dimer unit is strong enough, one electron (or hole) occupies a bonding (antibonding) MO in each dimer unit. 
This is identified as a Mott insulator, termed a dimer-Mott insulator, in the case in which the Coulombic interaction between electrons inside a dimer unit is larger than the bandwidth~\cite{kino}. 
Antiferromagnetic (AFM) alignment of electronic spins located at each dimer unit owing to the inter-dimer exchange interaction is a plausible magnetic structure (Fig.~\ref{fig:one}$\bm a$) that is often realized~\cite{jerome, dumm}. 

Now, a degree of freedom inside a dimer unit is taken into account. 
A charge degree of freedom inside a dimer unit, i.e. the shape of the electronic charge cloud, is activated in the case in which the inter-dimer couplings overcome the gap between the bonding and antibonding MOs. 
Inequivalent charge distribution in the two molecules in a dimer unit induces a local electric-dipole moment that is often called a ``dimer dipole". 
An alternate alignment of the dimer dipoles corresponding to the antiferroelectric (AFE)-type order is one possible configuration on the chain due to a gain of the interaction between electrons in the nearest neighboring dimers. 
We show here that this spin and charge configuration, termed the mutiferroic AFM+AFM order, shown in Fig.~\ref{fig:one}$\bm b$ has a key symmetry for the ME effects. 
This configuration is neither invariant by the space reversal operation (denoted by ${\mathcal I}_s$) nor by the time reversal operation (denoted by ${\mathcal I}_t$) as shown in Fig.~\ref{fig:one}$\bm c$, while it is invariant by the spontaneous space-time reversal operation (${\mathcal I}_s {\mathcal I}_t$).  
This symmetry consideration predicts a cross term of $P$ and $M$ in the Landau-type free energy giving rise to a linear ME effect. 

The above prediction from the symmetry consideration is embodied by the following microscopic picture. 
Let us set up the multiferroic AFE+AFM state in a chain, in which local magnetic moments are not fully polarized due to thermal and/or quantum fluctuations.
Spins are assumed to be directed, for example along an axis perpendicular to the chain, due to a weak anisotropic interaction, such as the Dzyaloshinskii-Moriya interaction.  
When electric field is applied along the chain, as shown in Fig.~\ref{fig:one}$\bm d$,  the electronic charge distributions in a right-polarized and left-polarized dimers are no longer symmetric to each other under the space reversal operation, making the exchange interactions acting on the up- and down-spin moments inequivalent to each other. 
As a result, the up- and down-spin polarizations are not canceled perfectly and a net magnetization appears. 
This is the spin-electronic contribution in the ME polarizability~\cite{malashevich}. 

\begin{figure} 
\begin{center}
    \includegraphics[width=0.9\columnwidth,clip]{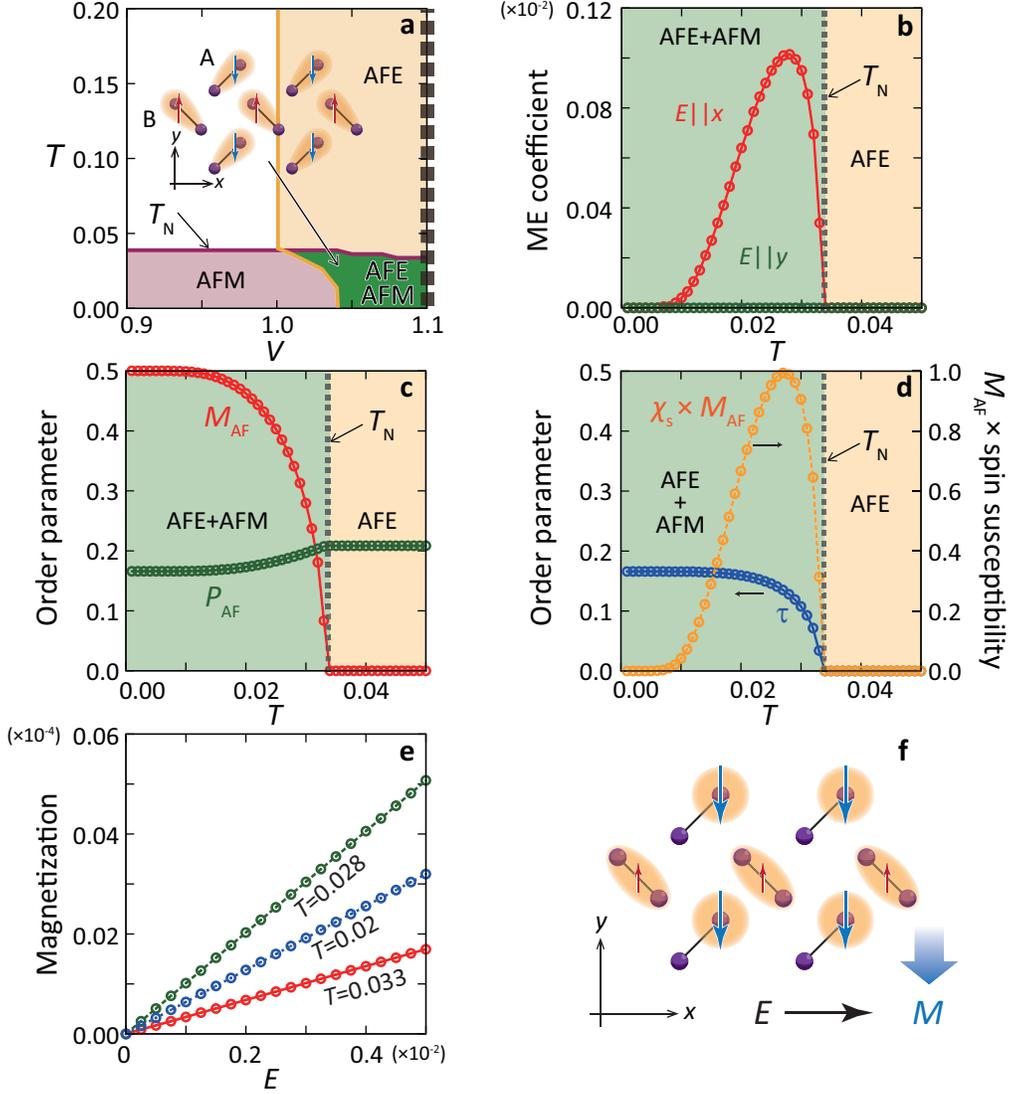}
\end{center}
\caption
{
{\bf ME effects in a two dimensional molecular dimer system.} 
${\bm a}$, 
Finite $T$ phase digram. 
Spin and charge configuration in the mutiferroic AFE+AFM phase is also shown. 
Filled purple circles, shaded ovals, and thin red and blue arrows represent molecules, charge distributions, and spins, respectively. 
The two kinds of dimers are labeled by A and B. 
${\bm b}$, Calculated $T$ dependences of the ME coefficients $\alpha$.  
Red and green lines represent the results in the case where the electric field is applied along the $x$ and $y$ axes, respectively.  
${\bm c}$, Calculated $T$ dependences of  the staggered magnetization $M_{\rm AF}$ (red line) and staggered electric polarization $P_{\rm AF}$ (green line). 
${\bm d}$, Calculated $T$ dependences of the composite spin-charge order parameter $\tau$ (blue line) and a product of $M_{\rm AF}$ and the magnetic susceptibility $\chi_s$ (orange line). 
${\bm e}$, Net magnetization versus electric field for several $T$. 
${\bm f}$, A spin and charge configuration below $T_{\rm N}$ under the electric field along the $x$ axis. 
}
\label{fig:kappa}
\end{figure}

While the essence of the ME effect in the molecular dimer system is incorporated in this simple one-dimensional model, in reality such quantum spin chains often show the spin-Peierls states at low temperatures, rather than the AFM order~\cite{yu, riera, kuwabara, yoshimi}. 
Thus, we demonstrate the ME effect by numerical calculations in a realistic $\kappa$-(BEDT-TTF) type crystal lattice shown in Fig.~\ref{fig:kappa}$\bm a$, where two-dimensional alignment of the dimer units prevents a paring of spins associated with the bond alternation. 
A minimal theoretical model for the molecular dimer systems showing the multiferroic AFE+AFM phase is known to be the Hamiltonian~\cite{naka, hotta} given by
\begin{eqnarray}
{\cal H}=
J\sum_{\langle ij \rangle} {\bm S}_i \cdot {\bm S}_j
-V\sum_{\langle ij \rangle} Q_i ^x Q_j^x
+\Gamma \sum_{i} Q_i^z  
-K\sum_{\langle ij \rangle} {\bm S}_i \cdot {\bm S}_j Q_i^x Q_j^x , 
\label{eq:hamiltonian}
\end{eqnarray}
where ${\bm S}_{i}$ is a spin operator at $i$-th dimer unit with magnitude of 1/2. 
The charge degree of freedom inside a dimer unit is represented by the pseudo spin operator, ${\bm Q}_i$, with magnitude of 1/2. 
The $x$ component, $Q^x_i=+1/2 \ (-1/2)$, represents an electrically polarized state, and the $z$ component, $Q^z_i=+1/2 \ (-1/2) $, represents a bonding (antibonding) MO state, where an electronic charge distribution is symmetric in a dimer unit. 
All interaction parameters are positive. 
The first term represents the conventional AFM Heisenberg interaction. 
The second and third terms, respectively, originate from the inter-dimer Coulomb interaction and the electron hopping between MOs inside the dimer units, and promotes and prevents the long-range order of the dimer dipoles. 
The last term represents a coupling between spins and dimer dipoles, and has a similarity to the Kugel-Khomskii  type Hamiltonian for the orbital degenerated transition-metal compounds~\cite{kugel}. 
This model is derived from the generalized Hubbard-type model by the perturbational calculations, that are presented in the {\it Supplemental Information} (SI).
In the following, $\Gamma/2$ corresponding to the intra-dimer hopping integral is taken as a unit of energy, which is approximately 0.3 eV for the typical $\kappa$-type BEDT-TTF compounds. 

A phase diagram on a plane of temperature ($T$) and the inter-dimer Coulomb interaction ($V$) calculated by the mean-field approximation introduced in SI is presented in Fig.~\ref{fig:kappa}$\bm a$. 
In low temperatures, the two typical phases are confirmed; 
a multiferroic AFE+AFM ordered phase, where spins align antiferromagnetically and polarizations of the electronic clouds induce the canted AFE dimer-dipole order, as shown Fig.~\ref{fig:kappa}$\bm a$, and an electrically non-polarized phase with AFM order, where the electronic clouds distribute symmetrically inside the dimer units. 
Spins are assumed to be directed along the $y$ axis due to a weak anisotropic interaction which is not included in the model explicitly. 
We focus on the multiferroic AFE+AFM ordered phase.  
Realizations of this phase were suggested experimentally in the BEDT-TTF compounds~\cite{lang, iguchi}. 

A magnetization, an electric polarization, and ME response coefficients are calculated in finite $T$. 
An ME response coefficient, $\alpha_{\mu \nu}=dM_{\mu}/dE_{\nu} |_{E=0}$, is presented in Fig.~\ref{fig:kappa}$\bm b$. 
In the parameter sets adopted in the numerical calculations (the dotted line in Fig.~\ref{fig:kappa}$\bm a$), the AFE-type dipole order, characterized by an order parameter $P_{\rm AF}$, occurs at a much higher temperature than the N$\rm \acute e$el temperature ($T_{\rm N}$), as presented in Fig.~\ref{fig:kappa}$\bm c$. 
It is shown that $\alpha_{\mu \nu}$ emerges below $T_{\rm N}$ and disappears toward the zero temperature. 
This temperature dependence almost traces the magnetic fluctuation, shown in Fig.~\ref{fig:kappa}$\bm d$, where a product of the magnetic susceptibility ($\chi_s$) and the AFM order parameter ($M_{\rm AF}$) is plotted, indicating that the fluctuation is responsible for the ME effect as mentioned above. 
A large anisotropy in the tensor components of $\alpha$ is seen; there is no ME response when $E$ is parallel to the $y$ axis, because the electronic clouds for the up- and down-spins are equivalent even under the electric field. 
That is, $\alpha_{yx}$ is only finite in these spin and charge configurations. 
Although the ordered spins are assumed to be directed along the $y$ axis in the present calculation, the tensor components of $\alpha_{\mu \nu}$ emerge in a similar manner in the case where spins in the AFM phase are directed along other directions. 
A linearlity of the induced $M$ with respect to $E$ expected from the symmetry consideration is obtained as shown in Fig.~\ref{fig:kappa}$\bm e$. 
A schematic spin and charge configuration under the electric field applied to the $x$ axis is shown in Fig.~\ref{fig:kappa}$\bm f$. 
We have checked that the inverse ME response coefficient, ${\bar \alpha}_{\mu \nu}=dP_{\mu}/dH_{\nu}|_{H=0}$, shows the same temperature dependence with $\alpha$. 
Since the magnetic field perpendicular to the spin direction does not break an equivalence of the two kinds of the polarized dimers, ${\bar \alpha}_{xy}$ is only finite. 

\begin{figure} 
\begin{center}
   \includegraphics[width=0.9\columnwidth,clip]{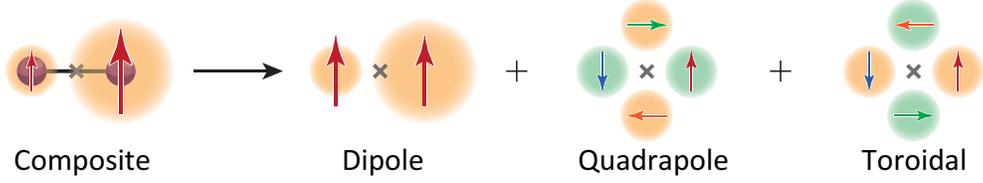}
\end{center}
\caption
{
{\bf Spin and charge distributions for the spin-charge composite moment.} 
Filled circles, shaded ovals, and thin arrows represent molecules, charge distributions, and spins, respectively. 
Shaded orange and green circles represent positive and negative charge distributions, respectively.
The spin distribution in a dimer unit in the ferroic spin-charge composite ordered state is decomposed into a magnetic dipole moment, a magnetic quadrapole moment, and a toroidal moment. 
}
\label{fig:toroidal}
\end{figure}

So far, our discussion of the ME effect has been restricted in the multiferroic AFE+AFM ordered state, where the equivalence of the up- and down-spin sublattices or that of the two kinds of dimer-dipole sublattices is broken by the external fields. 
The necessary condition of the ME effect is generalized by introducing the composite operator of the spin and charge degrees of freedom defined by $\bm{\mathcal T}_i=\sum_{\nu} 2 p_i^{\nu} {\bm S}_i$, termed the spin-charge composite operator, in which ${\bm p}$ is a local electric dipole moment at $i$-th dimer. 
A local dipole moment $\bm p$ is represented by $p_{i}^{x} = p_{i}^{y} = Q_{i}^{x}$ for the A dimers and $- p_{i}^{x} = p_{i}^{y} = Q_{i}^{x}$ for the B dimers, respectively, where the A and B dimers are defined in Fig. 2(a). 
From the viewpoint of the multipole moment, the spin distribution of this object is reduced to the magnetic dipole, magnetic quadrapole, and toroidal moments, as shown in Fig.~\ref{fig:toroidal}, and the charge distribution is reduced to the electric dipole and electric quadrapole moments. 
This operator changes its sign by the space reversal operation, ${\cal I}_s$, as well as by the time reversal one, ${\cal I}_t$, but it is invariant by the simultaneous operation of ${\cal I}_s$ and ${\cal I}_t$.  
We show in the following that the ferroic order of $\bm{\mathcal T}_i$, i.e., $\tau = N^{-1} \left| \sum_i \langle \bm{\mathcal T}_i \rangle \right|$ with the number of dimers $N$, gives rise to the ME effect, even with neither the AFM order nor the AFE order. 
Temperature dependence of $\tau$ in the case of the multiferroic AFE+AFM state introduced above is presented in Fig.~\ref{fig:kappa}$\bm d$. 

In order to demonstrate this concept of the spin-charge composite order, we set up a model for the molecular dimer system where quenched randomness is introduced. 
This is modeled by randomly directed local electric field, $h_i$, acting on the dimer dipoles. 
This is introduced as $\sum_{i} h_i Q_i^x$, in addition to the Hamiltonian defined in equation~(\ref{eq:hamiltonian}). 
Possible origins of this term in the BEDT-TTF compounds are attributed to the random configurations of the ethylene groups in the BEDT-TTF molecules~\cite{muller, su} and the random orientations of the CN groups in the anion layer~\cite{tomic}.  
Artificial X-ray irradiation may also produce random potentials in samples~\cite{sasaki, sasaki2}.
Relaxor-like behaviors in the dielectric constant, which might be due to random dipole configurations, are often observed experimentally in the dimer-type organic molecular solids~\cite{jawad, lang, iguchi, jawad2}. 
The model Hamiltonian with the random electric field is analyzed by the cluster mean-field approximation, in which physical quantities are averaged with respect to the random configurations of $h_i$, and amplitude of the random field is denoted by $h$. 
Details are given in SI. 

\begin{figure} 
\begin{center}
   \includegraphics[width=0.9\columnwidth,clip]{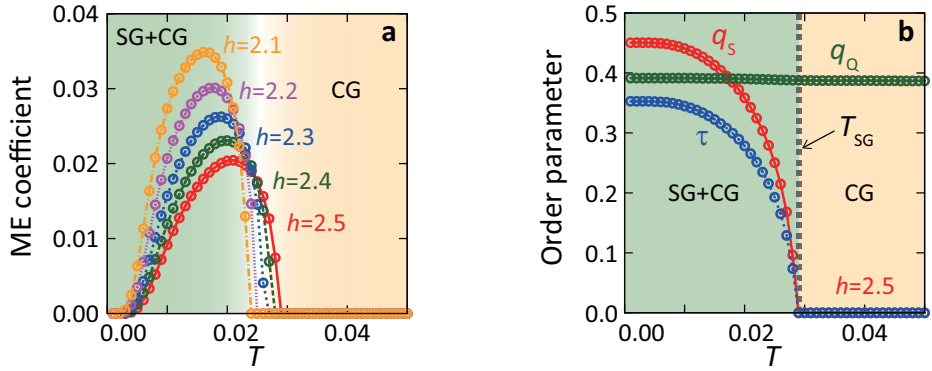}
\end{center}  
\caption
{
{\bf ME effects in a two dimensional molecular dimer system in the presence of local random electric fields.}
${\bm a}$, Calculated $T$ dependences of the ME coefficient for the several amplitudes of the random field  $h$. 
The electric field is applied along to the $y$ direction. 
${\bm b}$, 
Calculated $T$ dependences of order parameters at the random field amplitude $h = 2.5$. 
Order parameters for SG ($q_{S}$) and  CG ($q_{Q}$) are denoted by red and green lines, respectively. 
A blue line represents the ferroic spin-charge composite order parameter. 
}
\label{fig:random}
\end{figure}

As shown Fig.~\ref{fig:random}$\bm a$, finite values of the ME coefficients emerge below certain temperatures. 
In the case of the strong randomness, instead of the multiferroic AFE+AFM state, the spin glass (SG) state associated with the electric-dipole glass, i.e. the charge glass (CG) state emerges. 
In Fig.~\ref{fig:random}$\bm b$, we plot the temperature dependences of the SG order parameter ($q_S$) and the CG order parameter ($q_Q$).
While the CG order parameter is always finite, the SG state sets in at a certain temperature denoted by $T_{\rm SG}$. 
Any types of order parameters for the conventional magnetic and electric-dipole long-range orders are zero in a whole temperature range, unlike the case without the randomness. 
There is a hidden order below $T_{\rm SG}$, i.e., the ferroic order of the spin-charge composite operator appears, as shown in Fig.~\ref{fig:random}$\bm b$. 
It is shown in Fig.~\ref{fig:random}$\bm a$ that the linear ME coefficients for several amplitudes of the randomness appear in concert with $\tau$. 
The present ME effect is active even without the conventional magnetic and electric-dipole orders, but under the ferroic order of the spin-charge composite object. 

The present scenario for the ME effect has significant potentialities for actual dimer-type organic molecular solids. 
A possible candidate is $\kappa$-(BEDT-TTF)$_2$Cu[N(CN)$_2$]Cl. 
A long-range order of the dimer dipoles associated with the AFM order, which is similar to the configuration shown in Fig.~\ref{fig:kappa}$\bm a$, was reported below the N$\rm \acute e$el temperature at approximately 27 K~\cite{lang}, 
although there is a debate for a realization of the dimer dipoles~\cite{dressel}. 
Another candidate is $\beta'$-(BEDT-TTF)$_2$ICl$_2$ where a change in the dielectric responses was observed at the  N$\rm \acute e$el  temperature~\cite{iguchi}. 
The present scenario of the ME effect is also applicable to a series of TMTTF$_2X$; in the case of $X=$SbF$_6$, a ferroelectric-type dipole order associated with the AFM order emerges~\cite{chow, brazovskii, monceau}. 
The expected maximum values of the ME and inverse ME coefficients from the present theory are of the order on $10^{-6}$--$10^{-4}$ in the cgs Gauss system, in which $10^{-4} $ is the same order of the ME coefficients in Cr$_2$O$_3$~\cite{astrov}. 
The ME effect proposed here has a chance to be generalized into the ME effect in the high frequency region, which will be confirmed directly by the optical measurements. 
The present novel ME effect in the dimer-type organic molecular solids may not only provide the new guiding principle of multiferroic materials, but also promote material designs of organic molecular solids as flexible and lightweight multifunctional materials.

\bigskip

\noindent
{\bf Method}
\par \noindent
Phase diagram at finite temperature is calculated by applying the mean-field approximation to the Hamiltonian in equation~(\ref{eq:hamiltonian}), where $\langle S_i^{\mu}\rangle$, $\langle Q_i^{\mu}\rangle$, and $\langle S_i^{\mu} Q_i^{\nu}\rangle$ as the order parameters are determined self-consistently. 
The model Hamiltonian with the random field is analyzed by the cluster mean-field approximation; 
spin and pseudo spin states inside of small clusters with the mean field are calculated exactly. 
Expectation values are obtained by averaging in terms of the random field configurations.

\bigskip
\noindent
{\bf Acknowledgments}
\par
\noindent
We thank 
T.~Arima, J. Nasu and T. Watanabe for helpful discussions.
This work was supported in part by Core Research for Evolutional Science and Technology, Japan Science and Technology Agency, and Grant-in-Aid for Scientific Research Priority Area from the Ministry of Education, Science and Culture of Japan.
Parts of the numerical calculation was performed in the supercomputing facilities in Institute for Solid State Physics, the University of Tokyo. 

\bigskip
\noindent
{\bf Author contributions}
\par
\noindent
M.N. carried out the calculations. 
M.N. and S.I. analyzed the results. 
M. N. and S.I. wrote the paper. 
S.I. led the project. 

\bigskip
\noindent
{\bf Additional information}
\par
\noindent
Supplementary information is available in the online version of the paper. 
Reprints and permissions information is available online at www.nature.com/reprints. 
Correspondence and requests for materials should be addressed to S.I. 

\bigskip
\noindent
{\bf Competing financial interests}
\par
\noindent
The authors declare no competing financial interests. 

\clearpage

%
\clearpage
%
\begin{center}
\section*{\Large 
Supplemental Information of  \\``Magnetoelectric effect in organic molecular solids"
}
\end{center}
\section{Model Hamiltonian} 
\label{MH}
The model Hamiltonian for the molecular dimer systems introduced in equation~(1) in the main text is derived from the extended Hubbard-type Hamiltonian~\cite{naka, hotta}. 
A molecular dimer unit consisting of the two molecules termed $a$ and $b$ is introduced at each site in a crystal lattice. 
The Hamiltonian is given by 
${\cal H}_{\rm EH} = {\cal H}_{\rm intra} + {\cal H}_{\rm inter}$. 
The intra-dimer term is given by 
\begin{align}
{\cal H}_{\rm intra} = t_{A} \sum_{i \sigma} \left( c^{\dagger}_{i a \sigma} c_{i b \sigma}  + {\rm H. c.} \right) 
+ U \sum_{i \mu } n_{i \mu \uparrow} n_{i \mu \downarrow} + V_{A} \sum_{i} n_{i a} n_{i b}, 
\label{eq:intra}
\end{align}
in which $c_{i \mu \sigma}^\dagger$ ($c_{i \mu \sigma}$) is the creation (annihilation) operator for a hole with spin $\sigma$($=\uparrow, \downarrow$) and molecule $\mu$($=a, b$) at $i$-th dimer, 
and $n_{i \mu} (= \sum_{\sigma} n_{i \mu \sigma} = \sum_{\sigma} c^{\dagger}_{i \mu \sigma} c_{i \mu \sigma})$ is the number operator. 
We introduce the intra-dimer hopping integral ($t_A$), the electron-electron interaction inside a molecule ($U$), 
and the inter-molecule electron-electron interaction inside a dimer unit ($V_A$). 
The inter-dimer term is given by 
\begin{align}
{\cal H}_{\rm inter} &= {\cal H}_{t} + {\cal H}_{V} \nonumber \\
&= \sum_{\langle ij \rangle \mu \mu' \sigma} t^{\mu \mu'}_{ij} \left( c^{\dagger}_{i \mu \sigma} c_{i \mu' \sigma}  + {\rm H.c.} \right) 
+ \sum_{\langle ij \rangle \mu \mu'} V^{\mu \mu'}_{ij} n_{i \mu} n_{j \mu'} , 
\end{align}
where $t^{\mu \mu'}_{ij}$ and $V^{\mu \mu'}_{ij}$, respectively, are the hopping integral and the electron-electron interaction between the molecule $\mu$ at $i$-th dimer unit and the molecule $\mu'$ at $j$-th dimer unit.  

In the case of the strong electron-electron interaction inside the dimer, i.e. $U-t_{A}, V_{A}-t_{A} \gg t_{\mu \mu'}, \ V_{\mu \mu'}$, 
the number of hole at each dimer is fixed to be one, and the low-energy effective Hamiltonian is obtained by the perturbational calculations, in which  
${\cal H}_{\rm inter}$ is treated as the perturbational term. 
We introduce the pseudo spin (PS) operator ${\bm Q}$ with an amplitude $1/2$ to describe the charge degree of freedom inside the dimer defined by 
\begin{align}
{\bm Q}_{i} = \frac{1}{2} \sum_{\sigma \nu \nu'} {\hat c}^{\dagger}_{i \nu \sigma} {\bm \sigma}_{\nu \nu'} {\hat c}_{i \nu \sigma}, 
\end{align}
where we define the hole operators for the bonding and antibonding molecular orbitals as  
${\hat c}_{i \alpha(\beta) \sigma} =  (c_{ia\sigma} -(+) c_{ib\sigma})/\sqrt{2}$. 
The eigen functions of $Q^{x}$ are the charge polarized state where the $a$ or $b$ orbital is occupied by a hole, 
and those for $Q^{z}$ are the non-polarized states where the antibonding orbital $(\alpha)$ or the bonding orbital $(\beta)$ is occupied. 
The effective Hamiltonian up to the order of ${\cal O} ({\cal H}_{V})$ and ${\cal O} ({\cal H}^{2}_{t})$ is given by 
\begin{align}
{\cal H}_{\rm eff} = {\widetilde {\cal H}}_{\rm intra} + {\widetilde {\cal H}}_{V} + {\cal H}_{J}. 
\label{eq:heff}
\end{align}
The first two terms are represented by 
\begin{align}
{\widetilde {\cal H}}_{\rm intra} + {\widetilde {\cal H}}_{V} = \Gamma \sum_{i} Q^{z}_{i} + \sum_{\langle ij \rangle} W_{ij} Q^{x}_{i} Q^{x}_{j}, 
\label{eq:intra2}
\end{align}
where $\Gamma$($=2t_{A}$) is the intra-dimer electron transfer and $W_{ij}$($=V^{aa}_{ij} + V^{bb}_{ij} - V^{ab}_{ij} - V^{ba}_{ij} $) is the Coulomb interaction.  
The third term in equation~(\ref{eq:heff}) represents the exchange interaction originating from the second order perturbation with respect to ${\cal H}_{t}$, and is classified by the spin-singlet and spin-triplet intermediate states in the perturbational  processes as 
\begin{align}
{\cal H}_{J} = - \sum_{\langle ij \rangle} \left( \frac{3}{4} + {\bm S}_{i} \cdot {\bm S}_{j} \right) h^{T}_{ij} - \sum_{\langle ij \rangle} \left( \frac{1}{4} - {\bm S}_{i} \cdot {\bm S}_{j} \right) h^{S}_{ij} .  
\label{eq:ex}
\end{align}
Explicit forms of $h^{m}_{ij}$ ($m=T, S$) are represented by 
\begin{align}
h^{m}_{ij} = & \sum_{\nu_{1}, \nu_{2} = (\alpha, \beta)} J^{\nu_{1} \nu_{2}}_{mij} {\hat n}_{i \nu_{1}} {\hat n}_{j \nu_{2}} + \sum_{\gamma_{1}, \gamma_{2} = (+,-)} J^{\gamma_{1} \gamma_{2}}_{mij} Q^{\gamma_{1}}_{i} Q^{\gamma_{2}}_{j} \notag \\
& + \sum_{\nu = (\alpha, \beta)} \left ( J^{x \nu}_{mij} Q^{x}_{i} {\hat n}_{j \nu} + J^{\nu x}_{mij} {\hat n}_{i \nu} Q_{j}^{x} \right ) , 
\label{eq:hmij}
\end{align}
where $Q_{i}^{\pm} = Q^{x}_{i} \pm i Q^{y}_{i}$ and ${\hat n}_{i \nu} = \sum_{\sigma = (\uparrow, \downarrow)} {\hat c}^{\dagger}_{i \nu \sigma} {\hat c}_{i \nu \sigma}$. 
The exchange constants are given by 
$J^{\nu \nu}_{Tij} =({\hat t}^{2}_{\alpha \beta} + {\hat t}^{2}_{\beta \alpha})/\Delta_{\nu \nu}^{T} $, 
%
$ J^{\nu \bar{\nu}}_{Tij} = 
({\hat t}^{2}_{\alpha \alpha} + {\hat t}^{2}_{\beta \beta})/\Delta_{\nu \bar{\nu}}^{T}$, 
%
$J^{++}_{Tij} = J^{--}_{Tij} = - {\hat t}_{\alpha \beta} {\hat t}_{\beta \alpha} \left(\Delta_{\alpha \alpha}^{S -1} + \Delta_{\beta \beta}^{S -1}  \right ) $, 
%
$J^{+-}_{Tij} = J^{-+}_{Tij} = - 2 {\hat t}_{\alpha \alpha} {\hat t}_{\beta \beta}/\Delta^{T}_{\alpha \beta} $,
%
$J^{x \nu}_{Tij} = ({\hat t}_{\beta \beta} {\hat t}_{\alpha \beta} - {\hat t}_{\alpha \alpha} {\hat t}_{\beta \alpha}) 
\left( \Delta^{T -1}_{\nu \nu} +  \Delta^{T -1}_{\alpha \beta} \right) $, 
and 
%
$J^{\nu x}_{Tij} =J^{x \nu}_{Tij}  \left (  {\hat t}_{\nu \nu'}  \leftrightarrow  {\hat t}_{\nu' \nu}   \right ) $ 
for the spin-triplet intermediate states, 
and 
\begin{align}
J^{\nu \nu}_{Sij} = \frac{{\hat t}^{2}_{\alpha \beta} + {\hat t}^{2}_{\beta \alpha}}{ \Delta_{\nu \nu}^{S}} + 4 {\hat t}^{2}_{\nu \nu} 
\left( \frac{D^{2}_{\nu}}{ \Delta^{D_{+}}_{\nu \nu}} + \frac{D^{2}_{\bar{\nu}}}{ \Delta^{D_{-}}_{\nu \nu}} \right) ,
\end{align} 
\begin{align}
J^{\nu \bar{\nu}}_{Sij} = \frac{{\hat t}^{2}_{\alpha \alpha} + {\hat t}^{2}_{\beta \beta}}{\Delta_{\alpha \beta}^{S}} + 2 {\hat t}^{2}_{\nu \bar{\nu}} \left( \frac{1}{\Delta^{D_{+}}_{\alpha \beta}} + \frac{1}{ \Delta^{D_{-}}_{\alpha \beta}} \right) , 
\end{align}
\begin{align}
J^{++}_{Sij} = J^{--}_{Sij} = {\hat t}_{\alpha \beta} {\hat t}_{\beta \alpha} 
\left( \frac{1}{ \Delta_{\alpha \alpha}^{S}} + \frac{1}{ \Delta_{\beta \beta}^{S} }\right) 
+ 2 {\hat t}_{\alpha \alpha} {\hat t}_{\beta \beta} C_{+} C_{-} 
\left( \frac{1}{\Delta^{D+}_{\alpha \alpha}} + \frac{1}{\Delta^{D+}_{\beta \beta}} 
       - \frac{1}{\Delta^{D-}_{\alpha \alpha} }- \frac{1}{\Delta^{D-}_{\beta \beta} } \right) , 
\end{align}  
\begin{align}
J^{+-}_{Sij} = J^{-+}_{Sij} = \frac{2 {\hat t}_{\alpha \alpha} {\hat t}_{\beta \beta}}{\Delta_{\alpha \beta}^{S}} + 4 {\hat t}_{\alpha \beta} {\hat t}_{\beta \alpha} C_{+} C_{-} 
\left( \frac{1}{\Delta^{D+}_{\alpha \beta}} - \frac{1}{\Delta^{D-}_{\alpha \beta} } \right) , 
\end{align} 
\begin{align}
J^{x \nu}_{Sij} &= 
({\hat t}_{\beta \beta} {\hat t}_{\alpha \beta} + {\hat t}_{\alpha \alpha} {\hat t}_{\beta \alpha}) 
\left( \frac{1}{\Delta^{S}_{\nu \nu}} + \frac{1}{\Delta^{S}_{\alpha \beta}} \right)  \notag\\
&+ 2 {\hat t}_{\nu \nu} {\hat t}_{\bar{\nu} \nu} \left\{ D_{\nu} \left( D_{\nu} + D_{\bar{\nu}} \right) 
\left( \frac{1}{\Delta^{D_{+}}_{\nu \nu}} + \frac{1}{\Delta^{D_{+}}_{\alpha \beta}} \right) 
-  D_{\nu} \left( D_{\nu} - D_{\bar{\nu}} \right) \left( \frac{1}{\Delta^{D_{-}}_{\nu \nu}} + \frac{1}{\Delta^{D_{-}}_{\alpha \beta}} \right) \right\} , 
\end{align}
and 
$J^{\nu x}_{Sij} = J^{x \nu }_{Sij} \left ( {\hat t}_{\nu \nu'}  \leftrightarrow {\hat t}_{\nu' \nu} \right ) $
for the spin-singlet intermediate states.
We define $\bar{\nu} = (\beta, \alpha)$ for $\nu = (\alpha, \beta)$, and 
${\hat t}_{\nu \nu'} \equiv {\hat t}^{\nu \nu'}_{ij} = \sum_{\mu, \mu' = (a, b)} U_{\nu \mu} t^{\mu \mu'}_{ij} U^{\dagger}_{\mu' \nu'}$. 
Energy differences are defined by  $\Delta^{T}_{\nu \nu'} = E_{T} - E_{\nu} - E_{\nu'}$,  $\Delta^{S}_{\nu \nu'} = E_{S} - E_{\nu} - E_{\nu'}$, and $\Delta^{D_{\pm}}_{\nu \nu'} = E_{D_{\pm}} - E_{\nu} - E_{\nu'}$ with $E_{T}=V_A$,  $E_{S}=U$, $E_{D_\pm}=(U + V_{A})/2 \pm \sqrt{4 t^{2}_{A} + (U - V_{A})^{2}/4}$, $E_{\alpha}=-t_A$, and $E_{\beta}=t_A$. 
Coefficients are $D_{\nu(\bar{\nu})} = \left\{ C_{+(-)}, C_{-(+)} \right\}$ for $\nu = (\alpha, \beta)$, 
where $C^{2}_{+} + C^{2}_{-} = 1$ and $C_{-}/C_{+} = \left\{ 2E_{D+} + 4t_{A} - (U + V_{A}) \right\}/(U - V_{A})$ are satisfied.

\begin{figure} 
\begin{center}
\includegraphics[width=0.9\columnwidth,clip]{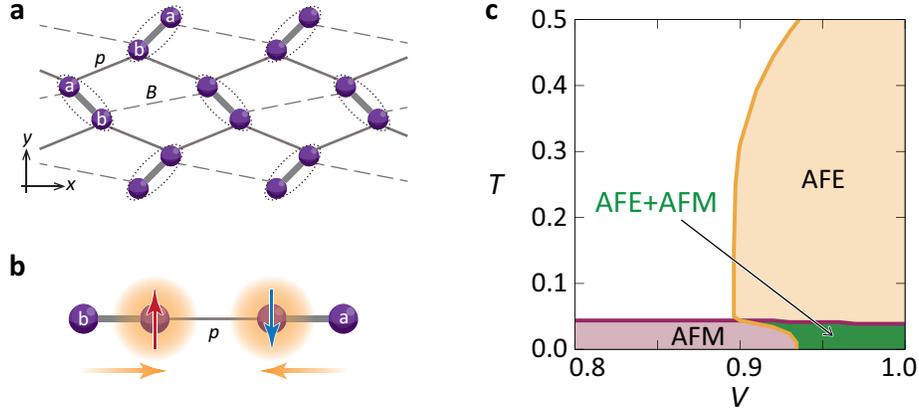}
\end{center}
\caption
{
{\bf Lattice structure of the $\kappa$-type BEDT-TTF compounds and the dominant exchange interaction.} 
${\bm a}$, A schematic lattice structure where filled circles denoted by $a$ and $b$ represent the BEDT-TTF molecules. 
Symbols $B$ and $p$ are the names of the bonds. 
Dotted ovals denote the dimer units. 
${\bm b}$, 
A stable spin and charge configuration in the nearest neighboring two dimer units connected by the $p$ bond. 
Filled circles, shaded circles, thin arrows, and shaded arrows indicate
molecules, charge distributions, spins, and electric dipoles, respectively.
${\bm c}$, 
A phase diagram calculated based on the Hamiltonian in equation~(\ref{eq:heff}), where all of the exchange interaction terms are taken into account. 
}
\label{fig:klatt}
\end{figure}
We apply this model to a two-dimensional plane in the crystal lattice of the $\kappa$-type BEDT-TTF compounds shown in Fig.~\ref{fig:klatt}$\bm a$. 
We focus on the ME effect in the AFE+AFM ordered phase. 
The calculated results are shown in Fig.~2 in the main text. 
Analyses of the ME effect without the conventional magnetic long-range orders will be discussed in Sect.~\ref{sect:randm}, and the results are shown in Fig.~4 in the main text. 
In the numerical calculations, 
parameter values are chosen to be $U=6$ and $V_{A}=4.5$, where $t_{A}$ is taken as a unit of energy. 
As for the inter-dimer hopping integral, the most dominant term on a bond denoted by $p$ shown in Fig.~\ref{fig:klatt}$\bm a$ is taken into account. 
A numerical value of the hopping integral is chosen to be  $t_{p} = 0.5$. 
This simplification might correspond to the $\kappa$-type BEDT-TTF compounds showing the AFM order, for example $\kappa$-(BEDT-TTF)$_2$Cu[N(CN)$_2$]Cl, where the so-called magnetic frustration effect is weak~\cite{koretsune}. 

The dominant exchange interactions in addition to the unperturbed terms in equation~(\ref{eq:intra2}) are given by 
\begin{align}
{\cal H} = \Gamma \sum_{i} Q_{i}^{z} - V \sum_{\langle ij \rangle} Q^{x}_{i} Q^{x}_{j} + J \sum_{\langle ij \rangle} {\bm S}_{i} \cdot {\bm S}_{j} - K \sum_{\langle ij \rangle} {\bm S}_{i} \cdot {\bm S}_{j} Q^{x}_{i} Q^{x}_{j} ,  
\label{eq:simH}
\end{align} 
which is the Hamiltonian introduced in equation~(1) in the main text. 
The third and fourth terms describe the Heisenberg-type exchange interaction and the spin-charge coupling, respectively. 
The exchange constants are explicitly given by 
\begin{align}
J = \frac{1}{4} \sum_{\nu_1,\nu_2 =  (\alpha, \beta)} \left( J^{\nu_1\nu_2}_{Sij} - J^{\nu_1\nu_2}_{Tij} \right) 
\end{align}
and 
\begin{align}
K = \sum_{\gamma_1,\gamma_2 =  (+, -)} \left( J^{\gamma_1\gamma_2}_{Sij} - J^{\gamma_1\gamma_2}_{Tij} \right).
\end{align}
The third and fourth terms in equation~(\ref{eq:simH}) in an isolated $p$ bond favors the antiparallel alignments of spins and charge PSs, as shown in Fig.~\ref{fig:klatt}$\bm b$.
This originates from the kinetic energy gain of $t_{p}$. 
Amplitudes of $J$ and $K$ are about 2--10 times larger than other exchange constants. 
The phase diagram where all exchange interaction terms are taken into account is shown in Fig.~\ref{fig:klatt}$\bm c$. 
The phase diagram calculated based on the Hamiltonian in equation~(\ref{eq:simH}) 
shown in Fig.~2$\bm a$ in the main text well reproduces the result in Fig.~\ref{fig:klatt}$\bm c$, 
implying the relevance of the present Hamiltonian in equation~(\ref{eq:simH}). 
\vfill
\eject

\section{Mean-field approximation}
The Hamiltonian introduced in equation~(1) in the main text is analyzed by the mean-field (MF) approximation. 
We take the unit cell that includes the two nonequivalent dimers, and introduce the following MFs, 
$\langle S^{\mu} \rangle$, $\langle Q^{x} \rangle$, and $\langle S^{\mu}Q^{x} \rangle$ with $\mu = (x, y, z)$ in each dimer, where the bracket represents the thermal average. 
The interaction terms are decoupled as 
${\bm S}_{i} \cdot {\bm S}_{j} \rightarrow {\bm S}_{i} \cdot \langle {\bm S}_{j} \rangle + \langle {\bm S}_{i} \rangle \cdot {\bm S}_{j} - \langle {\bm S}_{i} \rangle \cdot \langle {\bm S}_{j} \rangle $, 
$Q^{x}_{i} Q^{x}_{j} \rightarrow Q^{x}_{i} \langle Q^{x}_{j} \rangle + \langle Q^{x}_{i} \rangle Q^{x}_{j} - \langle Q^{x}_{i} \rangle \langle Q^{x}_{j} \rangle $, and 
${\bm S}_{i} Q^{x}_{i} \cdot {\bm S}_{j} Q^{x}_{j} \rightarrow {\bm S}_{i} Q^{x}_{i} \cdot \langle {\bm S}_{j} Q^{x}_{j} \rangle + \langle {\bm S}_{i} Q^{x}_{i} \rangle \cdot {\bm S}_{j} Q^{x}_{j} - \langle {\bm S}_{i} Q^{x}_{i} \rangle \cdot \langle {\bm S}_{j} Q^{x}_{j} \rangle$. 
The MFs are determined selfconsistently. 
\vfill 
\eject

\section{Randomness and cluster mean-field approximation}
\label{sect:randm} 
Randomness is introduced as the random electric field, which couples to the dimer dipoles. 
This interaction is represented by 
\begin{align}
{\cal H}_{\rm r} = - \sum_{i} h_{i} Q^{x}_{i}, 
\label{eq:rand}
\end{align}
where  $h_{i}$ is the random electric field at the $i$-th dimer and is defined as $h_i=h \epsilon_i$ with  amplitude $h$ and the site-depend random variable $\epsilon_i=\pm 1$.  
This is determined by the bimodal distribution function given by 
\begin{align}
P(\epsilon_i) = \frac{1}{2} \left\{ \delta (\epsilon_i - 1) + \delta (\epsilon_i + 1) \right\} . 
\end{align}
An expectation value of an observable ${\cal A}$ is given by the configuration average defined by 
\begin{align}
[ \langle {\cal A} \rangle ] = \prod_{i} \int_{-\infty}^{\infty}  dh_{i} P(\epsilon_i) \langle {\cal A} \rangle(\{  h_{i} \}), 
\end{align}
where $\langle {\cal A} \rangle(\{  h_{i} \})$ is the thermal average for a certain random field configuration $\{ h_{i} \}$~\cite{nishimori}. 

A sum of the effective Hamiltonian introduced in equation~(\ref{eq:simH}) in Sect.~\ref{MH} and the random field term in equation~(\ref{eq:rand}) is adopted as the model Hamiltonian. 
We examine a possibility of the ME effects without conventional magnetic long-range orders, in contrast to the ME effect in the AFM ordered phase shown in Fig.~2 in the main text. 
Thus, we introduce the magnetic frustration effect, in addition to the randomness effects, and suppress development of the N$\acute{\rm e}$el order.  
Then, the exchange constants in equation~(\ref{eq:simH}) are estimated by considering the hopping integrals for the $B$ bonds, as well as those in the $p$ bonds (see Fig.~\ref{fig:klatt}$\bm a$). 
Numerical values of the hopping integrals are chosen to be $t_{B} = t_{p} = 0.5$. 
This might be suitable for the $\kappa$-type BEDT-TTF compounds, in which no conventional magnetic long-range order appears, for example $\kappa$-(BEDT-TTF)$_2$Cu$_2$(CN)$_3$, where the frustration effect is strong~\cite{koretsune}.

The Hamiltonian with the random electric field is analyzed by the cluster MF approximation, in which  the exact diagonalization methods based on the Householder algorithm and the MF approximation are combined. 
The MF decouplings are introduced as ${\bm S}_{i} \cdot {\bm S}_{j} \rightarrow {\bm S}_{i} \cdot \left[ \langle {\bm S}_{j} \rangle \right] + \left[ \langle {\bm S}_{i} \rangle \right] \cdot {\bm S}_{j} - \left[ \langle {\bm S}_{i} \rangle \right] \cdot \left[ \langle {\bm S}_{j} \rangle \right]$, 
$Q^{x}_{i} Q^{x}_{j} \rightarrow Q^{x}_{i} \left[ \langle Q^{x}_{j} \rangle \right] + \left[ \langle Q^{x}_{i} \rangle \right] Q^{x}_{j} - \left[ \langle Q^{x}_{i} \rangle \right] \left[ \langle Q^{x}_{j} \rangle \right]$, and 
${\bm S}_{i} Q^{x}_{i} \cdot {\bm S}_{j} Q^{x}_{j} \rightarrow {\bm S}_{i} Q^{x}_{i} \cdot \left[ \langle {\bm S}_{j} Q^{x}_{j} \rangle \right] + \left[ \langle {\bm S}_{i} Q^{x}_{i} \rangle \right] \cdot {\bm S}_{j} Q^{x}_{j} - \left[ \langle {\bm S}_{i} Q^{x}_{i} \rangle \right] \cdot \left[ \langle {\bm S}_{j} Q^{x}_{j} \rangle \right]$. 
A cluster including three dimer units with the periodic boundary condition is adopted. 
The spin-glass and charge-glass order parameters plotted in Fig.~4$\bm b$ in the main text are defined as 
\begin{align}
q_{S} = \sqrt{[ \langle S^{x} \rangle^2 ] + [ \langle S^{y} \rangle^2 ] + [ \langle S^{z} \rangle^2 ]} 
\end{align}
and 
\begin{align}
q_{Q} = \sqrt{[ \langle Q^{x} \rangle^2 ]},  
\end{align}
respectively.


\begin{thebibliography}{99}
\bibitem{curie} 
Curie, P. 
Sur la symetrie dans les phenomenes physiques, symetrie d'un champ electrique et d'un champ magnetique. 
{\it J. Phys.} {\bf 3} (Ser. III){\bf,} 393-415 (1894).
%
\bibitem{dzyaloshinskii}
Dzyaloshinskii, I. E. 
On the magneto-electrical effect in antiferromagnets. 
{\it Sov. Phys. JETP} {\bf 10}(3){\bf,} 628-629 (1960). 
%
\bibitem{astrov}
Astrov, D. N. 
Magnetoelectric effect in chromium oxide. 
{\it Sov. Phy. JETP} {\bf 13,} 729-733 (1961). 
%
\bibitem{folen}
Folen, V. J., Rado, G. T., and Stalder, E. W. 
Anisotropy of the magnetoelectric effect in Cr$_2$O$_3$. 
{\it Phys. Rev. Lett.} {\bf 6,} 607-608 (1961). 
%
\bibitem{smolenskii}
Smolenskii, G. A. and Chupis, I. E. 
Ferroelectromagnets. 
{\it Sov. Phys. Usp.} {\bf 25,} 475-493 (1982). 
%
\bibitem{fiebig}
Fiebig, M. 
Revival of the magnetoelectric effect. 
{\it J. Phys. D} {\bf 38,} R123-R152 (2005). 
%
\bibitem{cheong}
Cheong, S-W. and Mostovoy, M. 
Multiferroics: A magnetic twist for ferroelectricity. 
{\it Nature Mater.} {\bf 6,} 13-20 (2007). 
%
\bibitem{kimura}
Kimura, T. {\it et al.} 
Magnetic control of ferroelectric polarization. 
{\it Nature} {\bf 496,} 55-58 (2003). 
%
\bibitem{katsura}
Katsura, H., Nagaosa, N., Balatsky, A. V. 
Spin current and magnetoelectric effect in noncollinear magnets. 
{\it Phys. Rev. Lett.} {\bf 95,} 057205 (2005).
%
\bibitem{dagotto}
Sergienko, I.A. and Dagotto, E.  
Role of the Dzyaloshinskii-Moriya interaction in multiferroic perovskites. 
{\it Phys. Rev. B} {\bf 73,} 094434 (2006). 
%
\bibitem{mostovoy}
Mostovoy, M. 
Ferroelectricity in Spiral Magnets. 
{\it Phys. Rev. Lett.} {\bf 96,} 067601 (2006). 
%
\bibitem{kimura2}
Kimura, T., Lashley, J. C., and Ramirez, A. P. 
Inversion-symmetry breaking in the noncollinear magnetic phase of the triangular-lattice antiferromagnet CuFeO$_2$. 
{\it Phys. Rev. B} {\bf 73,} (R)220401 (2006). 
%
\bibitem{arima}
Arima, T. 
Ferroelectricity induced by proper-screw type magnetic order. 
{\it J. Phys. Soc. Jpn.} {\bf 76,} 073702 (2007). 
%
\bibitem{khomskii}
Khomskii, D. I. 
Electric dipoles on magnetic monopoles in spin ice. 
{\it Nature Comm.} {\bf 3,} 904 (2012).
%
\bibitem{srinivasan}
Srinivasan, G., Rasmussen, E. T., and Hayes, R. 
Magnetoelectric effects in ferrite-lead zirconate titanate layered composites: the influence of zinc substitution in ferrites. 
{\it Phys. Rev. B} {\bf 67,} 014418 (2003).
%
\bibitem{zheng}
Zheng, H. {\it et al.} 
Multiferroic BaTiO$_3$-CoFe$_2$O$_4$ nanostructures. 
{\it Science} {\bf 303,} 601 (2004). 
%
\bibitem{essin}
Essin, A. M., Moore, J. E., and Vanderbilt, D. 
Magnetoelectric polarizability and axion electrodynamics in crystalline insulators. 
{\it Phys. Rev. Lett.} {\bf 102,} 146805 (2009).
%
\bibitem{malashevich}
Malashevich, A.,  Coh, S., Souza, I., and Vanderbilt, D. 
Full magnetoelectric response of Cr$_2$O$_3$ from first principles. 
{\it Phys. Rev. B} {\bf 86,} 094430 (2012). 
%
\bibitem{fukuyama}
Seo, H., Hotta, C., and Fukuyama, H. 
Toward systematic understanding of diversity of electronic properties in low-dimensional molecular solids. 
{\it Chem. Rev.} {\bf 104,} 5005 (2004). 
%
\bibitem{kanoda}
Kanoda, K. 
{\it The Physics of Organic Superconductors and Conductors} 
Ch. 22 
(Springer, Berlin, 2008).
%
\bibitem{jawad}
Abdel-Jawad, M. {\it et al.} 
Anomalous dielectric response in the dimer Mott insulator $\kappa$-(BEDT-TTF)$_2$Cu$_2$(CN)$_3$. 
{\it Phys. Rev. B} {\bf 82,} 125119 (2010). 
%
\bibitem{lang}
Lunkenheimer, P. {\it et al.} 
Multiferroicity in an organic charge-transfer salt that is suggestive of electric-dipole-driven magnetism. 
{\it Nature Mater.} {\bf 11,} 755-758 (2012).
%
\bibitem{iguchi}
Iguchi, S. {\it et al.} 
Relaxor ferroelectricity induced by electron correlations in a molecular dimer Mott insulator. 
{\it Phys. Rev. B} {\bf 87,} 075107 (2013).
%
\bibitem{dressel}
Sedlmeier, K. {\it et al.}
Absence of charge order in the dimerized $\kappa$-phase BEDT-TTF salts. 
{\it Phys. Rev. B} {\bf 86,} 245103 (2012). 
%
\bibitem{yakushi}
Yakushi, K., Yamamoto, K., Yamamoto, T., Saito, Y., and Kawamoto, A. 
Raman spectroscopy study of charge fluctuation in the spin-liquid candidate $\kappa$-(BEDT-TTF)$_2$Cu$_2$(CN)$_3$. 
{\it J. Phys. Soc. Jpn.} {\bf 84,} 084711 (2015). 
%
\bibitem{kino}
Kino, H. and Fukuyama, H. 
Phase diagram of two-dimensional organic conductors: (BEDT-TTF)$_2$X. 
{\it J. Phys. Soc. Jpn.} {\bf 65,} 2158-2169 (1996). 
%
\bibitem{jerome}
J$\rm \acute e$rome, D. 
The physics of organic superconductivity. 
{\it Science} {\bf 252,} 1509-1514 (1991). 
%
\bibitem{dumm}
Dumm, M. {\it et al.} 
Electron spin resonance studies on the organic linear-chain compounds (TMTCF)$_2$X (C=S,Se; X=PF$_6$ ,AsF$_6$ ,ClO$_4$ ,Br). 
{\it Phys. Rev. B} {\bf 61,} 511 (2000). 
%
\bibitem{yu}
Yu, W. {\it et al.} 
Electron-lattice coupling and broken symmetries of the molecular salts (TMTTF)$_2$SbF$_6$. 
{\it Phys. Rev. B} {\bf 70,} (R)121101 (2004). 
%
\bibitem{riera}
Riera, J. and Poilblanc, D. 
Coexistence of charge-density waves, bond-order waves, and spin-density waves in quasi-one-dimensional charge-transfer salts. 
{\it Phys. Rev. B} {\bf 62,} (R)16243 (2000). 
%
\bibitem{kuwabara} 
Kuwabara, M., Seo, H., and Ogata, M. 
Coexistence of charge order and spin-Peierls lattice distortion in one-dimensional organic conductors. 
{\it J. Phys. Soc. Jpn.} {\bf 72,} 225-228 (2003).  
%
\bibitem{yoshimi}
Yoshimi, K., Seo, H., Ishibashi, S., and Brown, S. E. 
Tuning the magnetic dimensionality by charge ordering in the molecular TMTTF salts. 
{\it Phys. Rev. Lett.} {\bf 108,} 096402 (2012). 
%
\bibitem{naka}
Naka, M. and Ishihara, S. 
Electronic ferroelectricity in a dimer Mott insulator. 
{\it J. Phys. Soc. Jpn} {\bf 79,} 063707 (2010). 
%
\bibitem{hotta}
Hotta, C. 
Quantum electric dipoles in spin-liquid dimer Mott insulator $\kappa$-(ET)$_2$Cu$_2$(CN)$_3$. 
{\it Phys. Rev. B} {\bf 82,} (R)241104 (2010). 
%
\bibitem{kugel}
Kugel, K. I. and Khomskii, D. I. 
The Jahn-Teller effect and magnetism: transition metal compounds. 
{\it Sov. Phys. Usp.} {\bf 25,} 231-256 (1982). 
%
\bibitem{muller}
M\"{u}ller, J. {\it et al.} 
Evidence for structural and electronic instabilities at intermediate temperatures in $\kappa$-(BEDT-TTF)$_2$X for X=Cu[N(CN)$_2$]Cl, Cu[N(CN)$_2$]Br and Cu(NCS)$_2$: Implications for the phase diagram of these quasi-two-dimensional organic superconductors. 
{\it Phys. Rev. B} {\bf 65,} 144521 (2002). 
%
\bibitem{su}
Su, X.,  Zuo, F.,  Schlueter, J. A.,  Kelly, M. E., and Williams, J. M. 
Structural disorder and its effect on the superconducting transition temperature in the organic superconductor $\kappa$-(BEDT-TTF)$_2$Cu[N(CN)$_2$]Br. 
{\it Phys. Rev. B} {\bf 56,} (R)14056 (1998). 
%
\bibitem{tomic}
Pinteri$\rm \acute c$, M. {\it et al.}
What is the origin of anomalous dielectric response in 2D organic dimer Mott insulators $\kappa$-(BEDT-TTF)$_2$Cu[N(CN)$_2$]Cl and $\kappa$-(BEDT-TTF)$_2$Cu$_2$(CN)$_3$. 
{\it Physica B} {\bf 460,} 202-207 (2015). 
%
\bibitem{sasaki}
Sasaki, T., Oizumi, H., Yoneyama, N., Kobayashi, N., and Toyota, N. 
X-ray irradiation-induced carrier doping effects in organic dimer-Mott insulators. 
{\it J. Phys. Soc. Jpn.} {\bf 76,} 123701 (2007). 
%
\bibitem{sasaki2}
Sasaki, S., Iguchi, S., Yoneyama, N., and Sasaki, T. 
X-ray irradiation effect on the dielectric charge response in the dimer-Mott insulator $\kappa$-(BEDT-TTF)$_2$Cu$_2$(CN)$_3$. 
{\it J. Phys. Soc. Jpn.} {\bf 84,} 074709 (2015). 
%
\bibitem{jawad2}
Abdel-Jawad, M., Tajima, N., Kato, R., and Terasaki, I. 
Disordered conduction in single-crystalline dimer Mott compounds. 
{\it Phys. Rev. B} {\bf 88,} 075139 (2013).
%
\bibitem{chow}
Chow, D. S. {\it et al.} 
Charge ordering in the TMTTF family of molecular conductors. 
{\it Phys. Rev. Lett.} {\bf 85,} 1698 (2000). 
%
\bibitem{brazovskii}
Monceau, P., Nad, F. Y., and Brazovskii, S. 
Ferroelectric Mott-Hubbard phase of organic (TMTTF)$_2$X conductors. 
{\it Phys. Rev. Lett.} {\bf 86,} 4080 (2001). 
%
\bibitem{monceau}
Nad, F. and Monceau, P. 
Dielectric response of the charge ordered state in quasi-one-dimensional organic conductors. 
{\it J. Phys. Soc. Jpn.} {\bf 75,} 051005 (2006). 
\end{thebibliography}

\begin{thebibliography}{99}
\bibitem{naka}
Naka, M. and Ishihara, S. 
Electronic ferroelectricity in a dimer Mott insulator. 
{\it J. Phys. Soc. Jpn.} {\bf 79,} 063707 (2010). 
%
\bibitem{hotta}
Hotta, C. 
Quantum electric dipoles in spin-liquid dimer Mott insulator $\kappa$-(ET)$_2$Cu$_2$(CN)$_3$. 
{\it Phys. Rev. B} {\bf 82,} (R)241104 (2010). 
%
\bibitem{koretsune}
Koretsune, T. and Hotta, C. 
Evaluating model parameters of the $\kappa$- and $\beta'$-type Mott insulating organic solids. 
{\it Phys. Rev. B} {\bf 89,} 045102 (2014). 
%
\bibitem{nishimori}
Nishimori, H. and Oritiz, G. 
{\it  Elements of Phase Transitions and Critical Phenomena} 
(Oxford University Press, Oxford, 2011). 
\end{thebibliography}
\end{document}